\documentclass[11pt]{article}

\usepackage{jheppub}
\usepackage{diagbox,latexsym,amssymb,amsfonts,amsmath,slashed}
\usepackage{dsfont}

\usepackage{bm}

\newcommand{\be}{\begin{equation}}
\newcommand{\ee}{\end{equation}}
\newcommand{\M}{\mathcal{M}}

\newcommand{\Budapest}{E\"otv\"os University, Department of
  Theoretical Physics, P\'azm\'any P.\ s.\ 1/A, H-1117, Budapest,
  Hungary.}
\newcommand{\Frankfurt}{Institute for Theoretical Physics, Goethe Universit\"at Frankfurt, D-60438 Frankfurt am Main, Germany.}

\title{
Magnetized baryons and the QCD phase diagram: \\ NJL model meets the lattice
}

\author[a]{G.~Endr\H{o}di,}
\author[b]{G.~Mark\'o}

\affiliation[a]{\Frankfurt}
\affiliation[b]{\Budapest}

\emailAdd{endrodi@th.physik.uni-frankfurt.de}
\emailAdd{marko@achilles.elte.hu}

\abstract{
We determine the baryon spectrum of $1+1+1$-flavor QCD in the 
presence of strong background magnetic fields using lattice 
simulations at physical quark masses for the first time. 
Our results show a splitting within multiplets according to 
the electric charge of the baryons 
and reveal, in particular, a reduction of the 
nucleon masses for strong magnetic fields. 
This first-principles input is used to define 
constituent quark masses and is employed to 
set the free parameters of the Polyakov loop-extended 
Nambu-Jona-Lasinio (PNJL) model 
in a magnetic field-dependent manner. 
The so constructed model is shown to exhibit 
inverse magnetic catalysis at high temperatures and 
a reduction of the transition temperature as the magnetic 
field grows -- in line with non-perturbative lattice 
results.
This is contrary to the naive variant of this model, 
which gives incorrect results for this fundamental 
phase diagram.
Our findings demonstrate that the magnetic field dependence of the PNJL model can be 
reconciled with 
the lattice findings in a systematic way, employing 
solely zero-temperature first-principles input.
}

\begin{document}

\maketitle

\section{Introduction}
\label{sec:intro}

The impact of background electromagnetic fields on strongly interacting matter 
is relevant for a range of physical situations including off-central heavy-ion 
collisions, magnetized neutron stars and the evolution 
of the early universe~\cite{Kharzeev:2015znc}. 
In particular, the elementary properties of magnetized hadronic degrees of freedom are important for 
cold astrophysical environments. The masses of baryons and mesons 
enter the nuclear equation of state and influence
the mass-radius relations of magnetars. Together with hadronic decay rates, these also 
affect stability of such compact objects and cooling mechanisms
that characterize the emitted neutrino spectra~\cite{Giunti:2014ixa}. 
For heavy-ion collisions, the magnetic field is produced in the very early stages and 
is expected to be short-lived~\cite{Voronyuk:2011jd,Tuchin:2013ie}, primarily 
affecting heavy baryons. A special role might be played by charged vector mesons that were 
conjectured to condense for sufficiently strong magnetic fields~\cite{Chernodub:2010qx}.

Besides their phenomenological importance, magnetic fields also 
represent external probes of strongly interacting matter i.e.\ of 
the underlying theory, quantum chromodynamics (QCD). 
One particular feature of the magneto-response of QCD matter that 
received great attention in the last decade is the 
phase diagram for nonzero temperatures and static, spatially uniform 
background magnetic fields, see, e.g., the review~\cite{Andersen:2014xxa}.
This phase diagram features a chiral symmetry restoration/deconfinement 
crossover~\cite{Aoki:2006we,Bhattacharya:2014ara}, where the chiral condensate $\bar\psi\psi$ 
drops towards zero and, 
almost simultaneously, the Polyakov loop $P$ rises.
According to lattice simulations, the pseudo-critical
temperature $T_c$, where the transition occurs, is reduced\footnote{Early lattice simulations that observed an increase in $T_c(B)$ suffered 
from large lattice artefacts~\cite{DElia:2010abb}.} as the magnetic field strength $B$ 
grows~\cite{Bali:2011qj,Bali:2012zg,Endrodi:2015oba}.
For physical quark masses (i.e.\ such that the pion mass is $M_\pi=135\textmd{ MeV}$), this behavior 
emerges due to the non-trivial dependence 
of $\bar\psi\psi$ on the temperature and on the magnetic field. 
On the one hand, for temperatures well below $T_c$ the magnetic field enhances $\bar\psi\psi$ (a 
phenomenon referred to as magnetic 
catalysis~\cite{Shovkovy:2012zn}). On the other hand, for $T\approx T_c$ 
the opposite is observed and $\bar\psi\psi$ is reduced by $B$ (inverse magnetic 
catalysis~\cite{Bruckmann:2013oba}). 
While magnetic catalysis originates from the high degeneracy of the lowest Landau-level~\cite{Gusynin:1995nb,Bruckmann:2017pft},
inverse magnetic catalysis arises as a result of the rearrangement of gluonic configurations 
induced by the magnetic field -- it is thus a secondary effect that can be associated to 
the indirect interaction between $B$ and electrically neutral gluons via sea quark loops~\cite{Bruckmann:2013oba}.
This mechanism is suppressed if quarks are heavy. Indeed, 
contrary to the situation at the physical point, for sufficiently heavy quarks
($M_\pi\gtrsim 500\textmd{ MeV}$), inverse magnetic catalysis does not 
occur anymore~\cite{Endrodi:2019zrl} -- 
nevertheless, the transition temperature is still reduced by $B$~\cite{DElia:2018xwo}.

The above summarized results are based on first-principles lattice QCD simulations. 
Before these became available, a multitude of low-energy models and effective theories of QCD 
were also employed to investigate the phase diagram for $B>0$. 
Surprisingly, the initial studies~\cite{Fraga:2008qn} observed the exact opposite of the 
lattice results: magnetic catalysis at all temperatures and the 
enhancement of the transition temperature with growing $B$.
A prime example for this behavior was obtained in the Polyakov loop-extended 
Nambu-Jona-Lasinio (PNJL) model~\cite{Gatto:2010pt}, but various other models resulted 
in the same picture\footnote{It is worth mentioning that a decreasing transition temperature was 
observed in a few simple models~\cite{Fraga:2012fs,Fraga:2012ev}.}, see the reviews~\cite{Fraga:2012rr,Andersen:2014xxa}.
The failure of these approaches was associated to the fact that gluons 
merely enter as a static background in these models 
so that the indirect mechanism behind inverse magnetic 
catalysis cannot be truly captured.
Later it was recognized that including a $B$-dependence in model parameters 
might improve the situation and bring model calculations closer to the lattice results. While in the Polyakov loop-extended quark meson model,
this was shown to be insufficient to have a monotonically 
reducing $T_c(B)$~\cite{Fraga:2013ova}, other studies did 
profit from this strategy~\cite{Farias:2014eca,Ferreira:2014kpa,Ayala:2014iba,Ayala:2014gwa,Ferreira:2013tba,Braun:2014fua,Andersen:2014oaa,Mueller:2015fka,Avancini:2016fgq,Farias:2016gmy}.
In particular a PNJL model study~\cite{Ferreira:2014kpa}, 
this was performed by tuning the coupling $G(B)$ to reproduce the 
transition temperature $T_c(B)$ obtained on the lattice. While this shows that the model can 
be made compatible with full QCD, in this example the predictive power of the effective approach
is clearly lost. 

Let us emphasize that effective models, albeit approximations to full QCD, are helpful for identifying the relevant 
degrees of freedom and interaction mechanisms, and thus guide our understanding of the physics 
of strongly interacting matter in extreme environments. 
For large baryon chemical potentials, where lattice simulations 
are hindered by the sign problem, low-energy models represent one of the few possibilities 
for the investigation of the phase diagram. 
Therefore it is highly desirable to test the limitations of such models in cases where 
importance sampling-based lattice investigations can be performed -- like the phase diagram 
at nonzero magnetic field or at nonzero isospin density~\cite{Brandt:2017oyy}.

In the present paper our aim is to develop a systematic approach to fix the parameters of 
the PNJL model utilizing magnetic field-dependent, first-principles input {\it at zero temperature}\footnote{We note that the usual parameterization of the Polyakov loop potential, which we employ as well, relies on temperature dependent data, however the novel magnetic field dependent corrections we use are derived solely at vanishing temperature.}. In particular we 
determine the baryon spectrum in three-flavor QCD using continuum extrapolated 
lattice simulations with physical quark masses. From this analysis, $T=0$ constituent quark masses 
are inferred and used to set the model parameters in a magnetic field-dependent manner\footnote{The $B$-dependence of effective couplings was 
also the subject of Ref.~\cite{Braghin:2016zba, Braghin:2017zas}.}.
According to our results, the phase diagram of the so constructed 
{\it lattice-improved} PNJL model agrees
with all features of the available lattice findings. 
Our method may also be extended to further low-energy models of QCD. 
We note that a similar idea was pursued in Ref.~\cite{Aarts:2018glk}, where temperature-dependent baryon masses measured on the lattice~\cite{Aarts:2017rrl} were used in an improved hadron resonance gas model. The meson spectrum of NJL-type models was also the subject of lattice investigations~\cite{AliKhan:1993qk}.

Besides fixing free parameters of effective descriptions, our results 
constitute the first lattice determination of magnetized baryon 
masses at the physical point. This complements earlier lattice calculations 
of baryon masses with heavier-than-physical quarks~\cite{Martinelli:1982cb,Chang:2015qxa,Parreno:2016fwu}, 
meson masses~\cite{Bali:2011qj,Hidaka:2012mz,Bali:2017ian} and decay rates~\cite{Bali:2018sey}, 
and properties of heavy quarkonia~\cite{Bonati:2015dka} in strong 
magnetic fields.
Our results might provide useful information for magnetized compact stars and the early 
stages of heavy-ion collisions, as pointed out above.

This paper is structured as follows. In Sec.~\ref{sec:num_setup} we describe our numerical setup 
and measurement strategy and present the results for the baryon spectrum. 
This is followed by Sec.~\ref{sec:njl}, where the definition of the 
constituent quark masses and the details of our PNJL model are given. The results for the 
magnetic field-dependent model parameters and the thermodynamics of the model is presented
in Sec.~\ref{sec:results}. Finally in Sec.~\ref{sec:summary} we summarize our findings and 
give an outlook for potential future research.

\section{Baryon spectrum from lattice simulations}
\label{sec:num_setup}

Our numerical simulations are performed on $N_s^3\times N_t$ 
lattices with spacing $a$, using the tree-level Symanzik 
improved gauge action and three flavors ($u$, $d$ and $s$) 
of stout-improved rooted
staggered quarks~\cite{Aoki:2005vt}. The quark masses $m_u=m_d$ and $m_s$ are set to their physical values 
along the line of constant physics~\cite{Borsanyi:2010cj}. 
The magnetic field is chosen to point in the $z$ direction and is 
implemented via $\mathrm{U}(1)$ phases satisfying periodic 
boundary conditions~\cite{Bali:2011qj}. This setup gives rise to a quantized magnetic flux $N_b\in\mathds{Z}$, such that
\be
eB = 6\pi N_b \cdot (aN_s)^{-2}\,,
\ee
where $e>0$ is the elementary charge and the quark electric charges 
are set as $q_u=-2 q_d=-2q_s=2e/3$. 
The details of our lattice ensembles are listed in Refs.~\cite{Bali:2011qj,Bali:2012zg}. 

The baryon masses can be extracted from the exponential decay of baryon correlators $C_b(t)$
at large Euclidean times\footnote{We note that in the present study we do not aim for precision results 
for the magnetic moments (related to the weak magnetic field-region), 
but concentrate on strong magnetic fields, which will be relevant 
for the phase diagram, see below. Thus we do not consider spin-projected operators but look for the state that minimizes the baryon 
energy.}. We employ localized corner sources. To enhance statistics we average 
over sources living on different time-slices as well as at different spatial locations. 
In addition, a sum over spatial coordinates is performed at the sink (at $B=0$ summing over the 
$x$ and $y$ coordinate components achieves zero momentum projection $p_x=p_y=0$, 
while for $B>0$ it merely helps to reduce fluctuations~\cite{Bali:2017ian}).
For staggered quarks, single-time-slice baryon operators mix parity partners 
so that the correlator takes the form~\cite{Ishizuka:1993mt},
\be
\label{eq:correlator}
C_b(t)=
A\left[{\rm e}^{-\M_b t}
+(-1)^{t+1}\,{\rm e}^{-\M_b(N_t-t)}
)\right] 
+A'\left[(-1)^{t+1}\,{\rm e}^{-\M'_b t}+{\rm e}^{-\M'_b(N_t-t)}
\right]\,,
\ee
requiring a four-parameter fit to extract the mass of the baryon, $\M_b$, and of its parity partner, $\M'_b$. 
We consider members of the baryon octet, including baryons 
with strangeness $S=0$, $-1$ and $-2$.
In the effective model study we will only use four of the baryons $b=p,n,\Sigma^0$ and $\Sigma^+$, for reasons which will become clear later.

\begin{center}
\begin{figure}
\centerline{\includegraphics[width=0.5\textwidth]{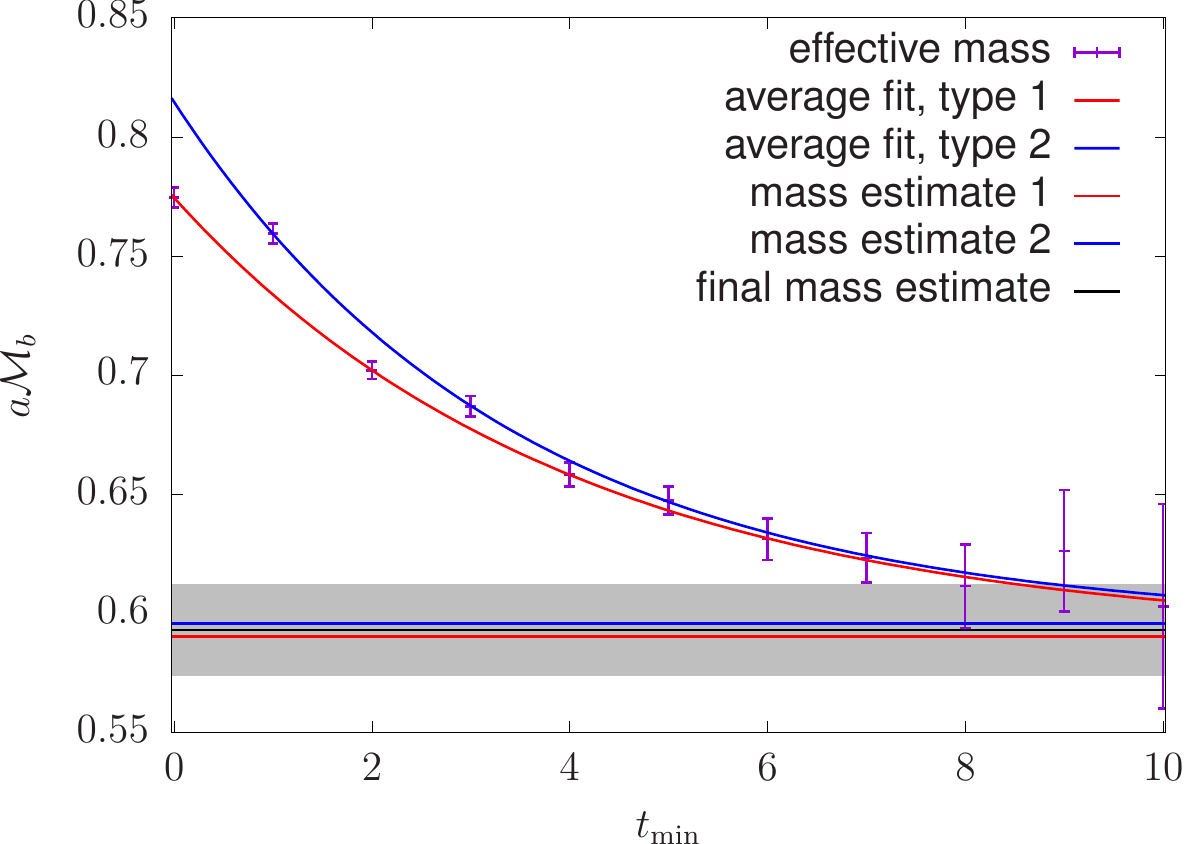}}\caption{\label{fig:effmass} Effective mass diagram of the proton at vanishing magnetic field. The parameter $t_{\rm min}$ characterizes the fitting region for \eqref{eq:correlator}: the larger it is, the more 
points are excluded from the fit. We estimate the $t_{\rm min}\to\infty$ limit by fitting exponential decays (type 1 -- only even points, type 2 -- only odd points), and the deviation of the estimates is used as the systematic error of our method. The grey band around the final mass estimate is the combined statistical and systematic error.}
\end{figure}
\end{center}

The determination of a baryon mass $\M_b$ at a certain lattice spacing $a$ and at a certain $B$ is as follows: an effective mass ($\M^{\rm eff}_b$) as a function of the fitting region (labeled by $t_{\rm min}$) is obtained by fitting the function \eqref{eq:correlator} to the measured correlator data in the region $[t_{\rm min},N_t-t_{\rm min}]$. A plateau is then extrapolated in $t_{\rm min}$ from the acquired $\M^{\rm eff}_b(t_{\rm min})$ as the $t_{\rm min}\to\infty$ limit of a simple exponential decay. The statistical error is then estimated with the jackknife method, while a systematic error is estimated from the exponential fit to find the plateau. The example of the proton effective mass is shown in Fig.~\ref{fig:effmass} along with the exponential fits and the mass estimates obtained, including statistical and systematic errors.

The continuum limit is carried out in two steps. First at vanishing magnetic field the masses $\M_b(B=0)$ are extrapolated to $a=0$, then separately only the magnetic field dependence $\M_b(B)/\M_b(B=0)$ is extrapolated to the continuum. The latter step requires interpolation for the magnetic field dependence, since at different lattice spacings we have measurements at different physical magnetic field values. We carry out the continuum limit of the $B$-dependence by fitting a Taylor-expansion with lattice spacing dependent coefficients,
\be
\label{eq:MB_extrap}
\frac{\M_b^2(eB,a)}{\M_b^2(0,a)} = 1+(c_0+c_1a^2)\cdot(eB)+(c_2+c_3a^2)\cdot(eB)^2+(c_4+c_5a^2)\cdot(eB)^3\,.
\ee
This ansatz is motivated by the $B$-dependence of the mass of a 
point-like charged particle.

\begin{center}
\begin{figure}
\centerline{\includegraphics[width=0.5\textwidth]{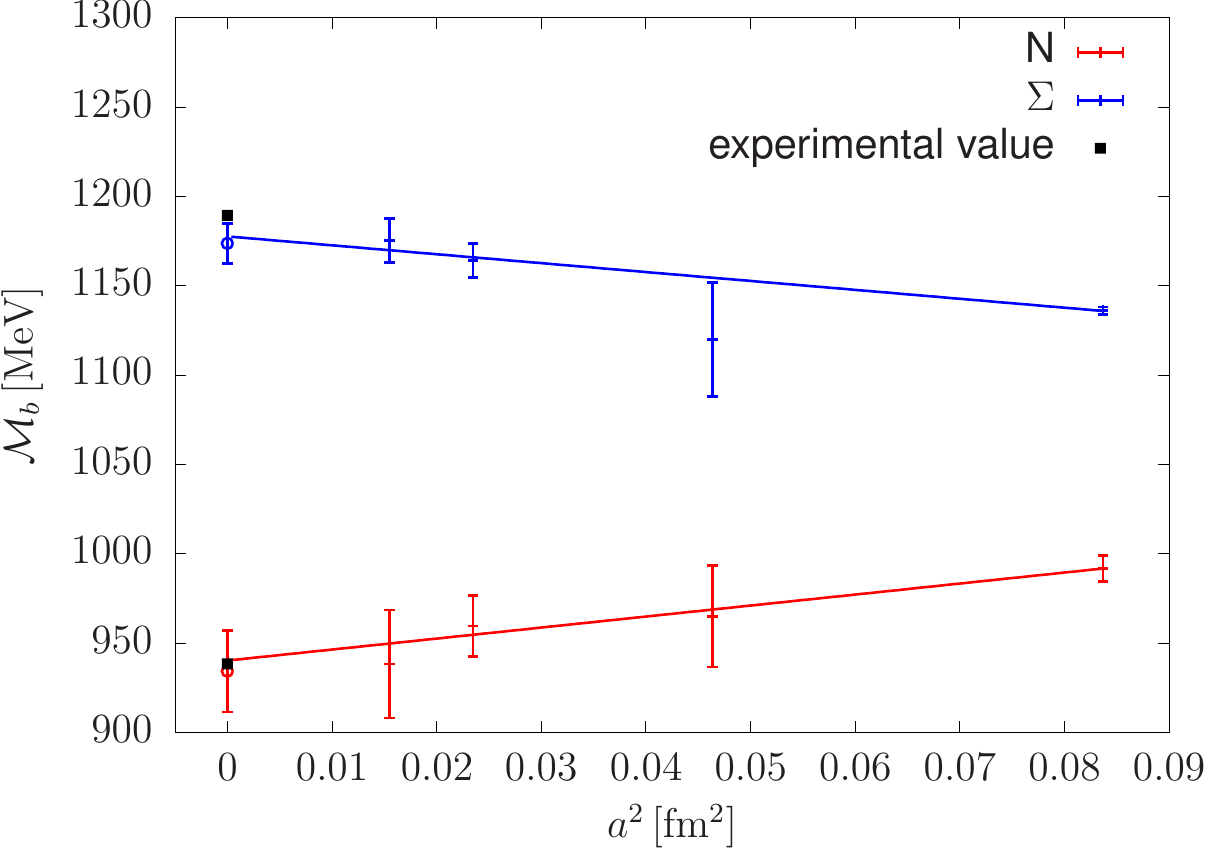}}\caption{\label{fig:B0contlim} Our continuum extrapolation of the masses of the nucleons and $\Sigma$ particles at $B=0$, where isospin symmetry is not yet broken. Also shown are the respective experimental values. The $\Xi$ particles need a more careful analysis due to their masses being distorted by close higher excitations and lattice artefacts therefore we disregard them here and in the rest of the paper.}
\end{figure}
\end{center}

\begin{center}
\begin{figure}
\includegraphics[width=\textwidth]{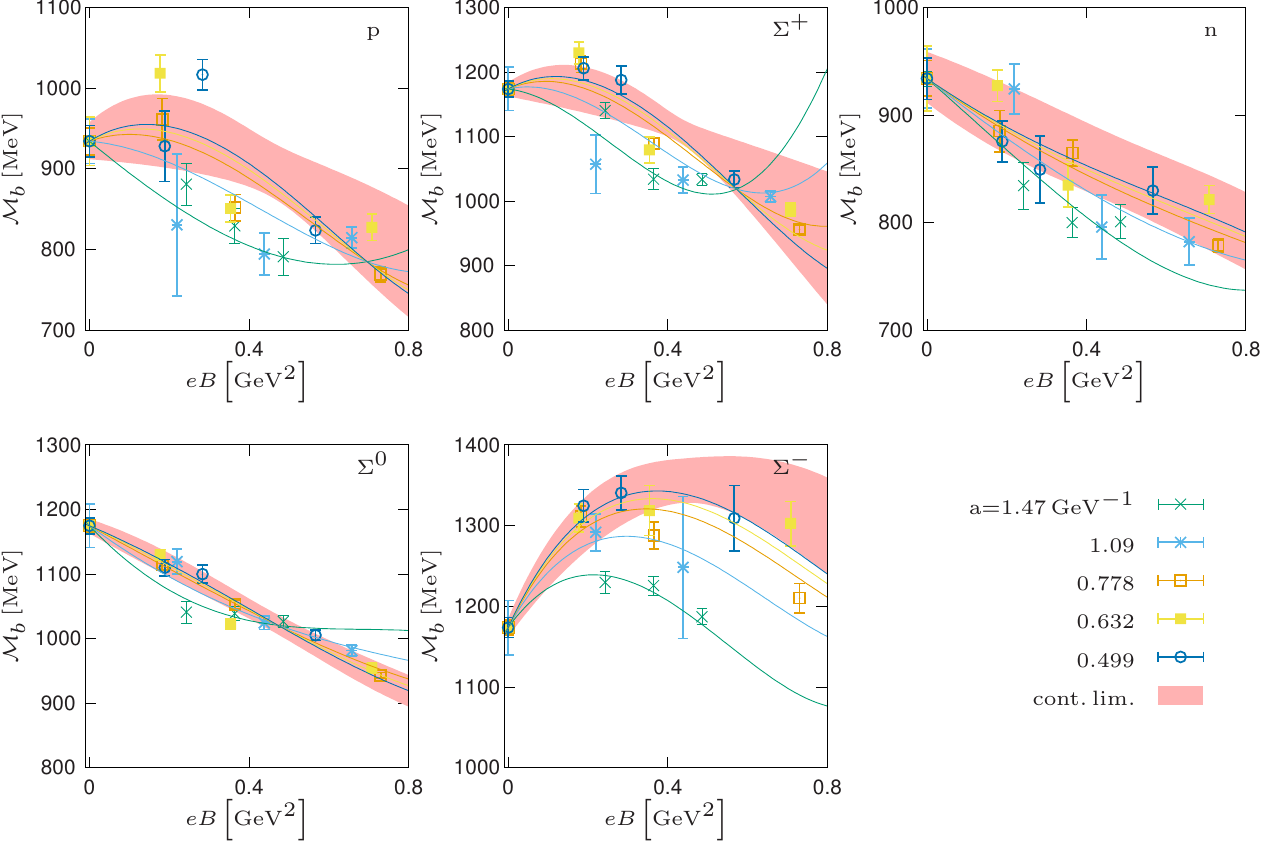}\caption{\label{fig:contlim_Bdep} Continuum extrapolation of the magnetic field dependent masses. The red bands are the estimates of the $\M_b(B)$ functions obtained by fitting the functional form \eqref{eq:MB_extrap} to the data points and evaluating at $a=0$. The colored lines show the sections of the fitted surface at the respective $a\neq0$ values.}
\end{figure}
\end{center}

To estimate the systematic error of our approach we redo the fits excluding the $B^3$ term to see how much the result changes. The statistical errors are estimated both for the $B=0$ and $B\neq0$ cases by the bootstrap method. The continuum extrapolation of the nucleon and $\Sigma$ masses at $B=0$ is shown in Fig.~\ref{fig:B0contlim}, comparing to their respective experimental values. 
Notice that at zero magnetic field isospin symmetry is present, which is reflected in our results as well. 
In the $S=-2$ channel, large lattice artefacts, together with the closeness of excited states prevent us from reaching an acceptable continuum limit
for the $\Xi$ baryons. 
(For precision results at $B=0$ 
including further baryons we refer the reader to Ref.~\cite{Durr:2008zz}.)
The continuum limit of the magnetic field dependence of the remaining baryon masses is shown in Fig.~\ref{fig:contlim_Bdep}. 
At low magnetic fields, a few outlier points are visible, 
related to the fact that here the Zeeman-splitting cannot be fully 
resolved. For strong magnetic fields this issue is absent. 
Notice furthermore that the behavior of the $\Sigma^-$ is completely different compared to the others. 
This might be explained within a simplified quark model: for all other baryons, quarks can orient their magnetic moments in an energetically favorable way with respect to the magnetic field in the lowest energy configuration (i.e.\ in the lowest Landau-level), however in the case of the $\Sigma^-$ one of the quarks is forced to be in an excited state (first Landau-level). 
Finally, a remark about the neutral $S=-1$ baryon (flavor content $uds$) is in order. Here we consider an operator that belongs to the octet at $B=0$, 
thus this particle is labeled as $\Sigma^0$. Nevertheless, at non-vanishing $B$, isospin is not a good quantum number anymore and $\Sigma^0$ mixes 
with the singlet state $\Lambda^0$, similarly to the mixing between 
$\rho$ and $\pi$ mesons~\cite{Bali:2017ian}. In this case we measure 
the lighter mixed state. Note furthermore that the $\Sigma^0$ and $\Lambda^0$ 
particles need not be distinguished for the definition of our 
constituent quark masses, see below.


\section{Construction of the PNJL model}
\label{sec:njl}
As an application for the $B$-dependent baryon masses, we use them as input for the magnetic field dependent reparameterization of the two-flavor PNJL model, which in turn will be used to explore the $B-T$ phase diagram of strongly interacting matter. First of all, since the PNJL model can only deal with constituent quark masses and not baryons, we use a simple non-relativistic quark model (NRQM) based on Ref.~\cite{Taya:2014nha} to define $B$-dependent $u,\,d$ and $s$ constituent quark masses. For this reason 
it is also advantageous to discuss baryons instead of mesons -- the 
latter receive their masses substantially from explicit chiral symmetry breaking and not from constituent quarks.

\begin{center}
\begin{figure}
\centerline{\includegraphics[width=0.5\textwidth]{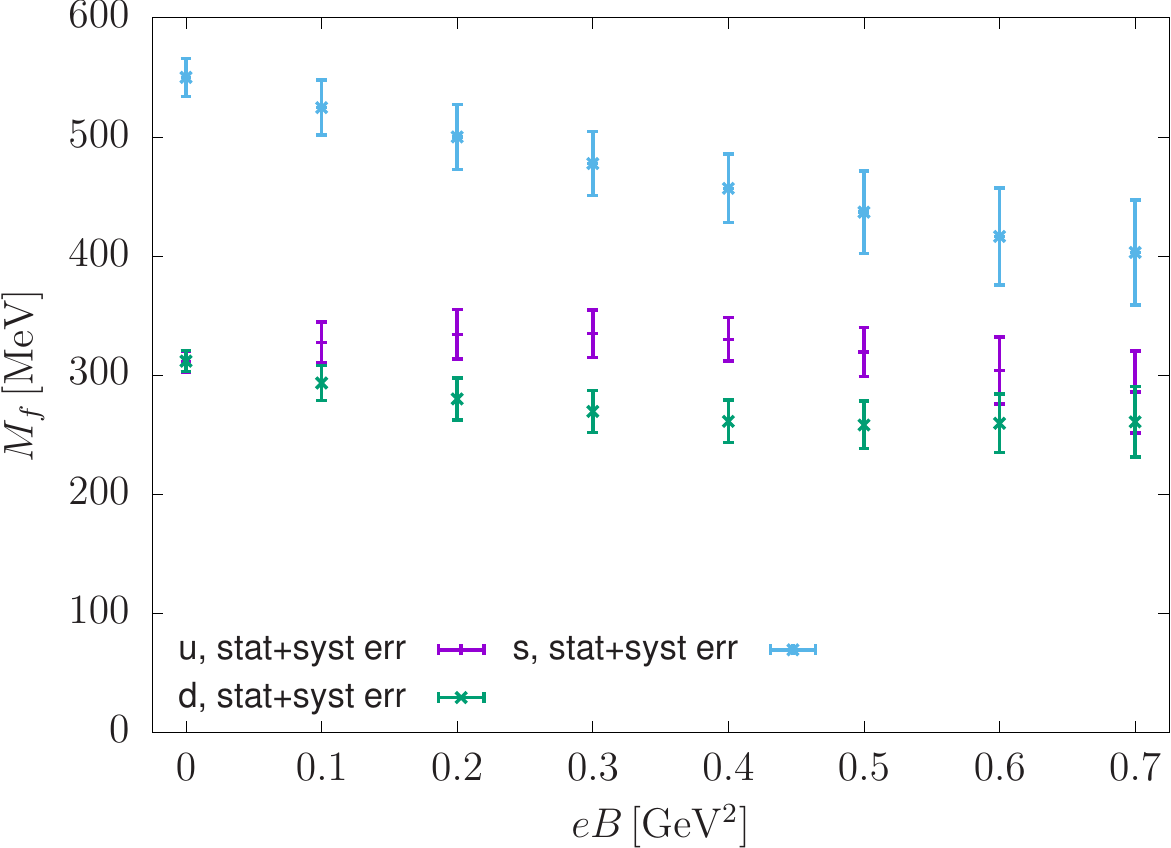}}\caption{\label{fig:cqms} Magnetic field dependent constituent quark masses as inferred form the continuum baryon masses shown in Fig.~\ref{fig:contlim_Bdep} using the NRQM based on Ref.~\cite{Taya:2014nha}. The systematic error is generally small compared to the statistical up to about $eB = 0.7$ GeV$^2$.}
\end{figure}
\end{center}

Our working assumption is that the baryon masses can be obtained by merely summing the masses of their constituents:
\be
\M_{b=\{f_1,f_2,f_3\}} = M_{f_1}+M_{f_2}+M_{f_3}\,,
\label{eq:qmodelansatz}
\ee
with $M_f$ being the constituent quark mass for flavor $f$. We determine $M_f$ as a function of $B$ by a least squares fit of the three quark masses to the results shown in Fig.~\ref{fig:contlim_Bdep}. According to Ref.~\cite{Taya:2014nha}, in the $\Sigma^-$ baryon at least one quark is forced into a spin state for which the Zeeman energy is added instead of subtracted. In the other four baryons, however, all quarks can be in the energetically most favorable spin state. To avoid having to describe excited states of the constituent quarks, we therefore disregard $\Sigma^-$ from the least squares fit. The goodness of the fits are found to be satisfactory, $\chi^2 < 1$ for all magnetic fields. The obtained constituent quark masses are shown in Fig.~\ref{fig:cqms}. The errors are propagated by bootstrap resampling, while systematic errors of the NRQM model are estimated by redoing the fits leaving out one baryon at a time.
We note that the simplistic ansatz~(\ref{eq:qmodelansatz}) could 
be improved by including the contribution of a 
$B$-dependent binding energy. Here we opted for including all $B$-dependence in the constituent quark masses, in order to make the connection with the the PNJL gap equation (where only $M_f$ enters)
more transparent.

We now briefly summarize the basic properties and equations of the PNJL model following Ref.~\cite{Fukushima:2010fe}, except that we use Schwinger's proper time method as the ultraviolet regularization scheme, see e.g. Ref.~\cite{Klevansky:1992qe}. Errors are propagated over from the constituent quark masses to all PNJL results by bootstrap resampling. The Lagrangian of the PNJL model is
\be
{\cal L} = \bar\psi(i\gamma_\mu D^\mu-m_0)\psi+G\left[(\bar\psi\psi)^2+(\bar\psi i\gamma_5\boldsymbol{\tau}\psi)^2\right]-{\cal U}(P,T)\,,
\ee
where $\psi$ is the constituent quark field coupled to the Polyakov loop $P$ through the covariant derivative and $m_0$ is the bare current quark mass. The Polyakov loop potential ${\cal U}(P,T)$ is a classical one constructed to reproduce pure gluonic lattice results for the temperature dependence of the Polyakov loop expectation value~\cite{Ratti:2005jh},
\be
{\cal U}(P,T) = T^4\left\{-\frac{a(T)P^2}{2}+b(T)\log\left[1-6P^2+8P^3-3P^4\right]\right\}\,,
\ee
with 
\be
a(T) = a_0+a_1\frac{T_0}{T}+a_2\left(\frac{T_0}{T}\right)^2\,\quad b(T) = b_3\left(\frac{T_0}{T}\right)^3\,.
\ee
We adopt the usual choice of parameters $a_0 = 3.51$, $a_1 = -2.47$, $a_2 = 15.2$, $b_3 = -1.75$, except for $T_0$, which sets the transition temperature in the pure gauge theory. It is usually set to $270$ MeV, however -- following Ref.~\cite{Schaefer:2007pw} -- we set it to $T_0=208$ MeV to include corrections induced by the two quark flavors. We use the mean-field approximation for the quarks, in which the thermodynamic potential at finite $B$ reads
\begin{align}
\Omega =\,\, &{\cal U}+\frac{(M-m_0)^2}{4G} \nonumber\\
&+\frac{T^2}{8\pi^2}\sum_{f=u,d}|q_fB|\overset{\;\;\;\infty}{\underset{T^2/\Lambda^2}{\int}}\frac{ds}{s^2}\coth\left(\frac{|q_fB|s}{T^2}\right){\,\rm e\,}^{\frac{-M^2s}{T^2}}\left[2\,\theta_3\left(\frac{\pi+\varphi}{2},{\,\rm e\,}^{-\frac{1}{4s}}\right)+\theta_3\left(\frac{\pi}{2},{\,\rm e\,}^{-\frac{1}{4s}}\right)\right]\,, \label{eq:pot}
\end{align}
where $M=M_u+M_d$ is the dynamically generated average constituent quark mass for the two flavors,
\be
\theta_3(p,q)\equiv\sum_{n=-\infty}^\infty q^{n^2}{\,\rm e\,}^{2inp}
\ee
is the third elliptic theta function and $\varphi$ marks the eigenvalue 
of the Polyakov loop matrix $L$ in the Polyakov gauge,
\be
L = \textmd{diag}(e^{i\varphi},e^{-i\varphi},1), \quad\quad
P=\frac{1}{3}\,\textmd{Tr} L = \frac{1}{3}(1+2\cos\varphi)\,.
\ee
In the mean-field approximation, both $M$ and $P$ minimizes the thermodynamic potential. We solve the model at every $B$ and $T$ by numerically searching for the two dimensional minimum of $\Omega$. Once the minimum is found, the quark condensate can be obtained by evaluating
\be
\label{eq:pbp}
\langle\bar\psi\psi\rangle=\frac{M}{4\pi^2}\sum_{f=u,d}|q_fB|\overset{\;\;\;\infty}{\underset{T^2/\Lambda^2}{\int}}\frac{ds}{s}\coth\left(\frac{|q_fB|s}{T^2}\right){\,\rm e\,}^{\frac{-M^2s}{T^2}}\left[2\,\theta_3\left(\frac{\pi+\varphi}{2},{\,\rm e\,}^{-\frac{1}{4s}}\right)+\theta_3\left(\frac{\pi}{2},{\,\rm e\,}^{-\frac{1}{4s}}\right)\right]\,.
\ee
Note that we follow the convention, where $\langle\bar\psi\psi\rangle$ is positive and therefore our \eqref{eq:pbp} contains an extra minus sign compared to most NJL studies.

\begin{center}
\begin{figure}
\centerline{\includegraphics[width=0.475\textwidth]{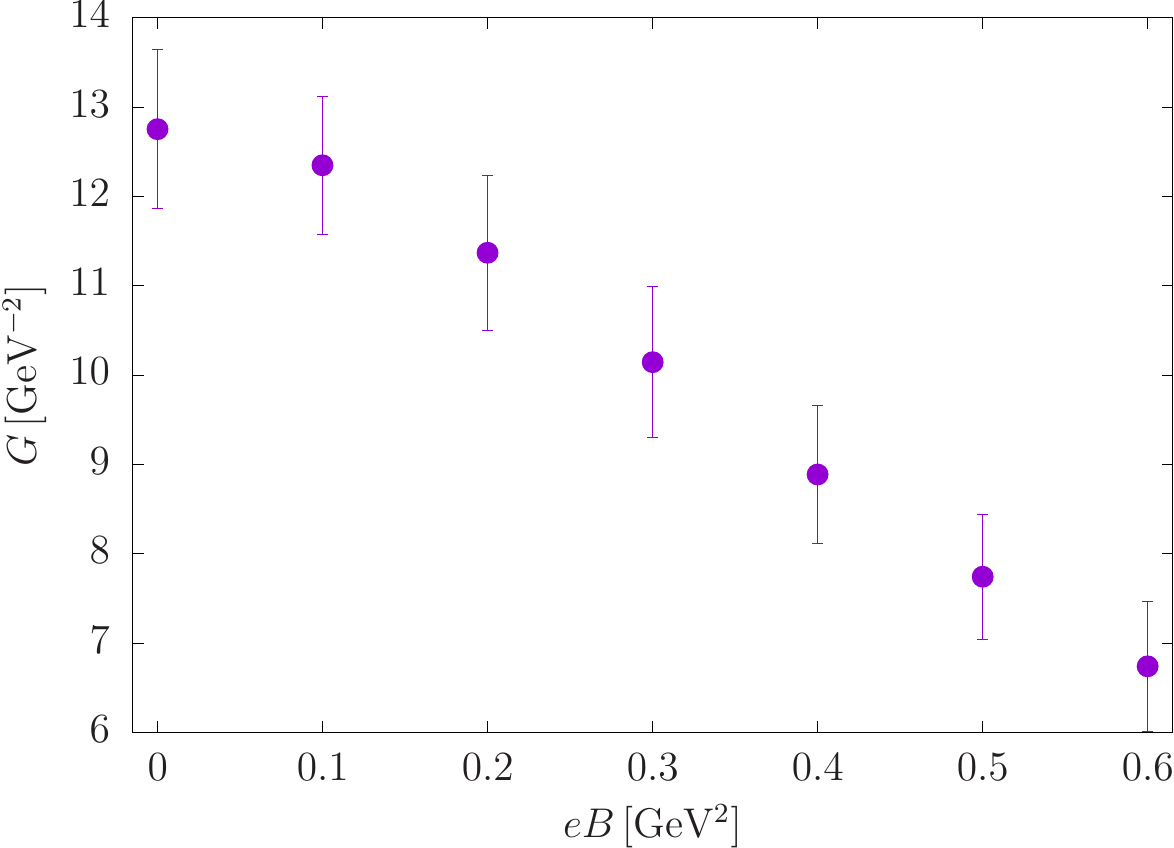}\hspace{0.05\textwidth}\includegraphics[width=0.475\textwidth]{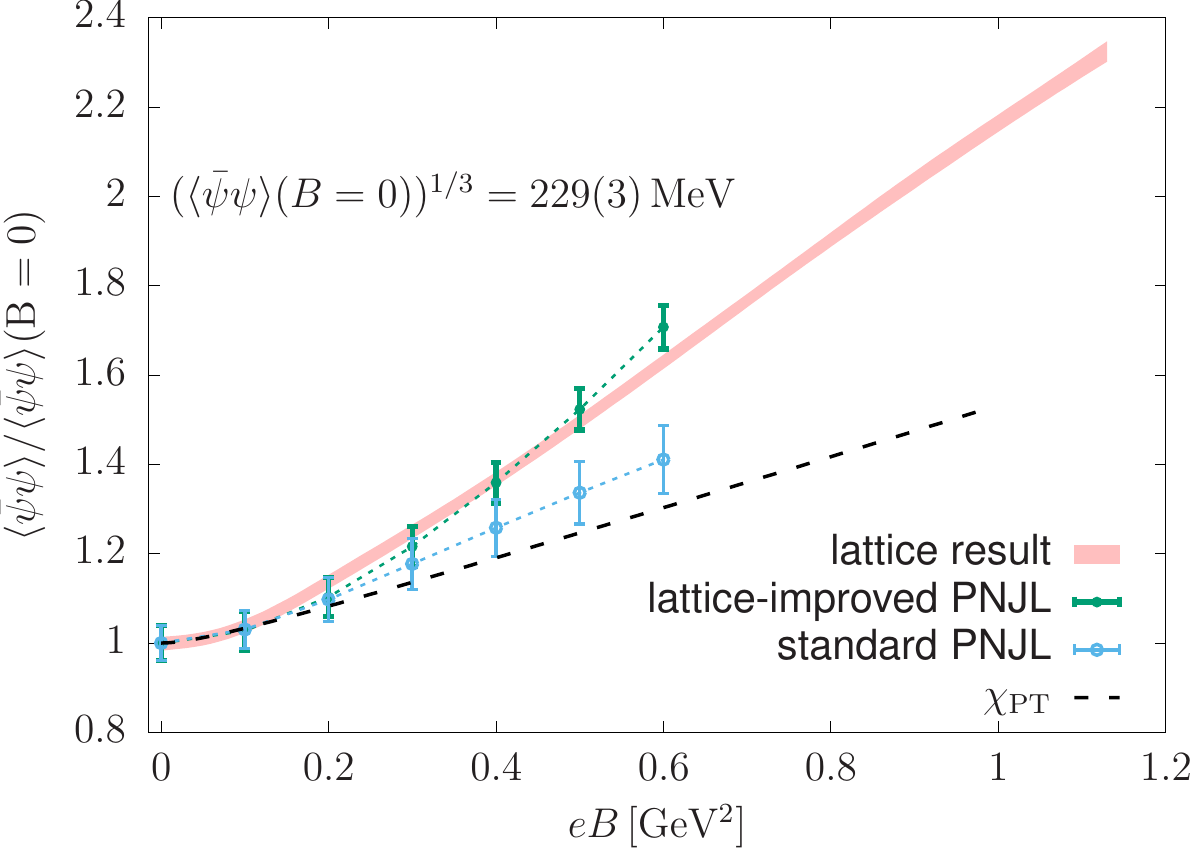}}\caption{\label{fig:GofB} Left: The magnetic field dependent coupling inferred from the baryon masses of Fig.~\ref{fig:contlim_Bdep}. The significant deviation from a constant already signals a strong effect on this level. Right: The average quark condensate at $T=0$ as a function of $B$ compared to lattice results of Ref.~\cite{Bali:2012zg}, to a standard PNJL calculation with $B$ independent coupling and to one-loop chiral perturbation theory \cite{Cohen:2007bt}.}
\end{figure}
\end{center}

The potential $\Omega$ depends on three model parameters: the bare current quark mass $m_0$, the four-fermion coupling $G$ and the cutoff scale $\Lambda$ of the theory (for different regularizations of the NJL model, see Ref.~\cite{Avancini:2019wed}). In mapping out the $B-T$ phase diagram we first fix $m_0$ and $\Lambda$ at $B=T=0$ by setting the predictions of the NJL model for the pion mass $m_\pi$,
\be
0 = 1-2G\Pi_{\rm \pi}(k^2=m_\pi^2) =
-\frac{6}{4\pi^2}\overset{\;\;\;\infty}{\underset{\Lambda^{-2}}{\int}}\frac{ds}{s^2}{\, \rm e \,}^{-M^2s} 
+\frac{6 m_\pi^2}{8\pi^2}\overset{\;\;\;\infty}{\underset{\Lambda^{-2}}{\int}}\overset{\;\;\;\infty}{\underset{\Lambda^{-2}}{\int}}\frac{ds_1ds_2}{(s_1+s_2)^2}{\, \rm e \,}^{-M^2(s_1+s_2)+\frac{s_1s_2m_\pi^2}{s_1+s_2}}\,,
\ee
and for the pion decay constant $f_\pi$,
\be
f_\pi^2=\overset{\;\;\;\infty}{\underset{\Lambda^{-2}}{\int}}\frac{ds}{s}{\, \rm e \,}^{-M^2s}
\ee
to their physical value, that is $138$ MeV and $93$ MeV, respectively, following Ref.~\cite{Klevansky:1992qe}. To fully fix the parameters of the NJL model we prescribe $M(B,T=0)$ to take the value which is consistent with the average of the $u$ and $d$ constituent quark masses inferred from the baryon masses measured on the lattice {\it for each} $B$. This results in $m_0 = 3.50(5)$ MeV and $\Lambda = 675(10)$ MeV and $G(B)$ plotted in Fig.~\ref{fig:GofB} (left) and listed together with the corresponding average constituent quark masses in Tab.~\ref{tab:GofB}.

\begin{table}
\centering
\begin{tabular}{|c|c|c|}
\hline
$eB$ [GeV$^2$] & $G$ [GeV$^{-2}$] & $M^2$ [GeV$^2$]\\\hline\hline
0.0 & 12.8(9) & 0.097(6)\\\hline
0.1 & 12.4(8) & 0.096(10)\\\hline
0.2 & 11.4(9) & 0.094(12)\\\hline
0.3 & 10.1(8) & 0.091(12)\\\hline
0.4 & 8.9(8) & 0.087(11)\\\hline
0.5 & 7.7(7) & 0.083(12)\\\hline
0.6 & 6.7(7) & 0.079(15)\\\hline
\end{tabular}
\caption{\label{tab:GofB} Values and errors of the magnetic field dependent four-fermion coupling and the average constituent quark masses used to fix them.}
\end{table}

We find that the coupling constant inferred from lattice baryon masses strongly decreases with increasing magnetic field. This reinforces studies which hand tuned the coupling to a qualitatively similar function in order to achieve the correct $T_c$ behavior. As a consistency check, in the right panel of Fig.~\ref{fig:GofB} we show a $T=0$ consequence of including the $B$-dependence in the coupling $G(B)$. The magnetic field dependence of the quark condensate is compared with lattice QCD results from Ref.~\cite{Bali:2012zg}. We see that our results are consistent with lattice results in a broad range of magnetic fields, while in a standard PNJL calculation where the coupling constant is a constant value (fixed to our $G(B=0)$) the two curves diverge for larger magnetic fields.

\section{Phase diagram}
\label{sec:results}
\begin{center}
\begin{figure}
\centerline{\includegraphics[width=0.475\textwidth]{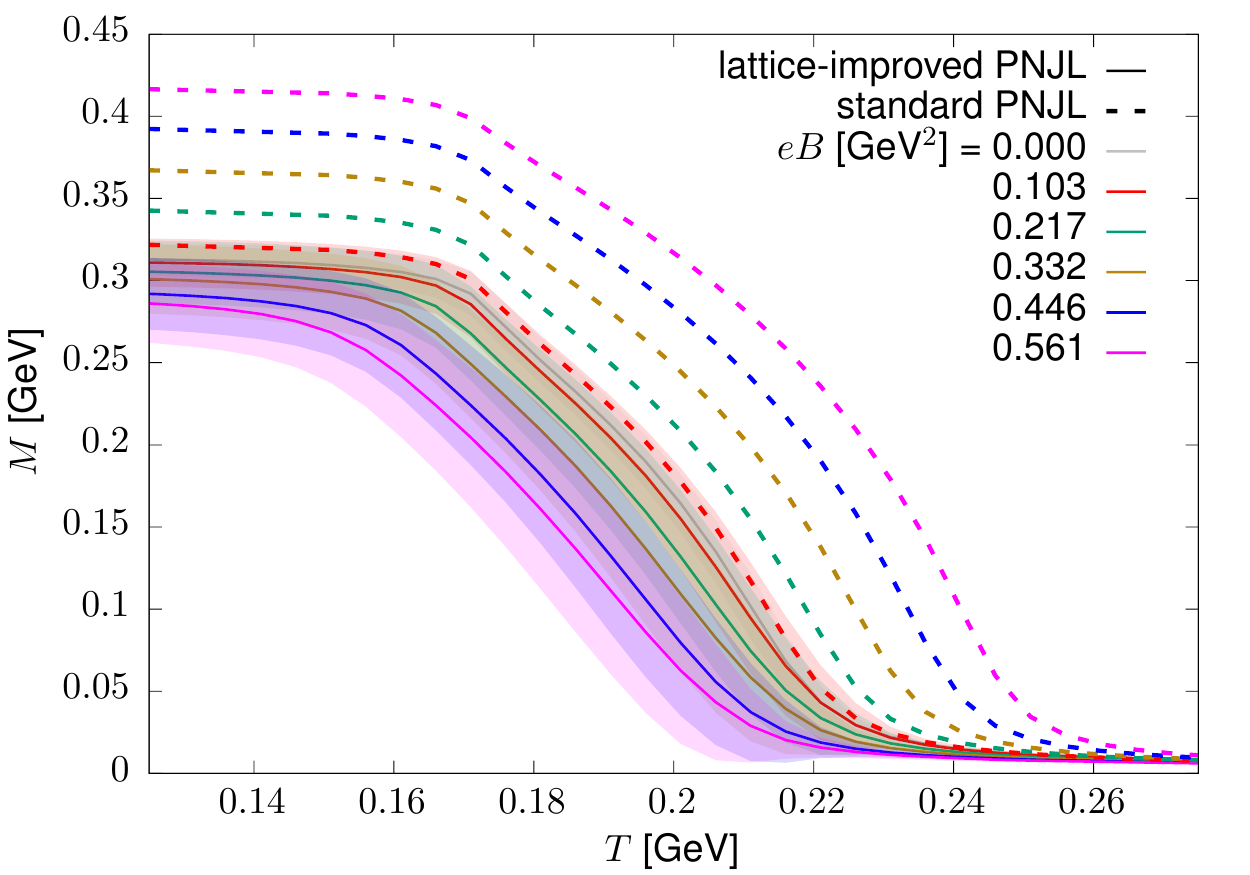}\hspace{0.05\textwidth}\includegraphics[width=0.475\textwidth]{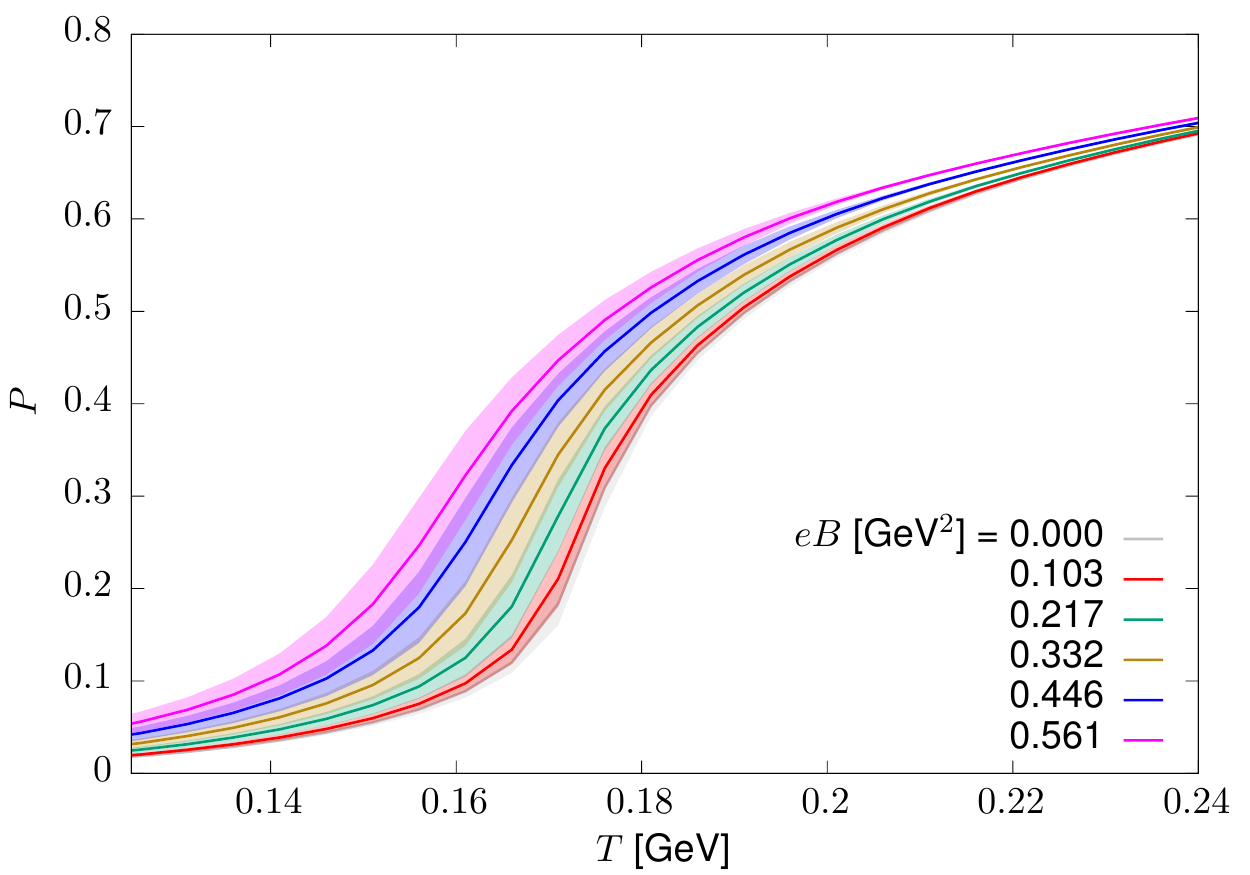}}\caption{\label{fig:njlsols} The solutions of the PNJL model utilizing the magnetic field-dependent coupling as functions of the temperature for different values of the magnetic field. Left: the average constituent quark mass. Also shown are standard PNJL results with $B$ independent coupling (errors are omitted on these curves for better visibility). Right: the expectation value of the Polyakov loop. In this case we do not show the results with $B$-independent coupling as they are indistinguishable from the grey $eB = 0$ curve.}
\end{figure}
\end{center}
We now turn to the results on mapping out the $B-T$ plane by minimizing the potential \eqref{eq:pot} with respect to both $M$ and $P$ using $G(B)$. The so obtained numerical solutions for $M(B,T)$ and $P(B,T)$ are shown in the left and right panels of Fig.~\ref{fig:njlsols} respectively. While at low temperature the dependence for different values of $B$ hardly changes, around the transition temperature larger $B$ leads to an earlier transition. The same cannot be seen in the standard PNJL results, where already at $T=0$ the mass grows significantly with $B$ and the transition is pushed further out as well. The Polyakov loop expectation values in the lattice-improved PNJL show the same behavior more pronounced, while in the standard PNJL we did not plot the Polyakov loop expectation value as it remains practically unchanged compared to the $B=0$ curve.

\begin{center}
\begin{figure}
\centerline{\includegraphics[width=0.475\textwidth]{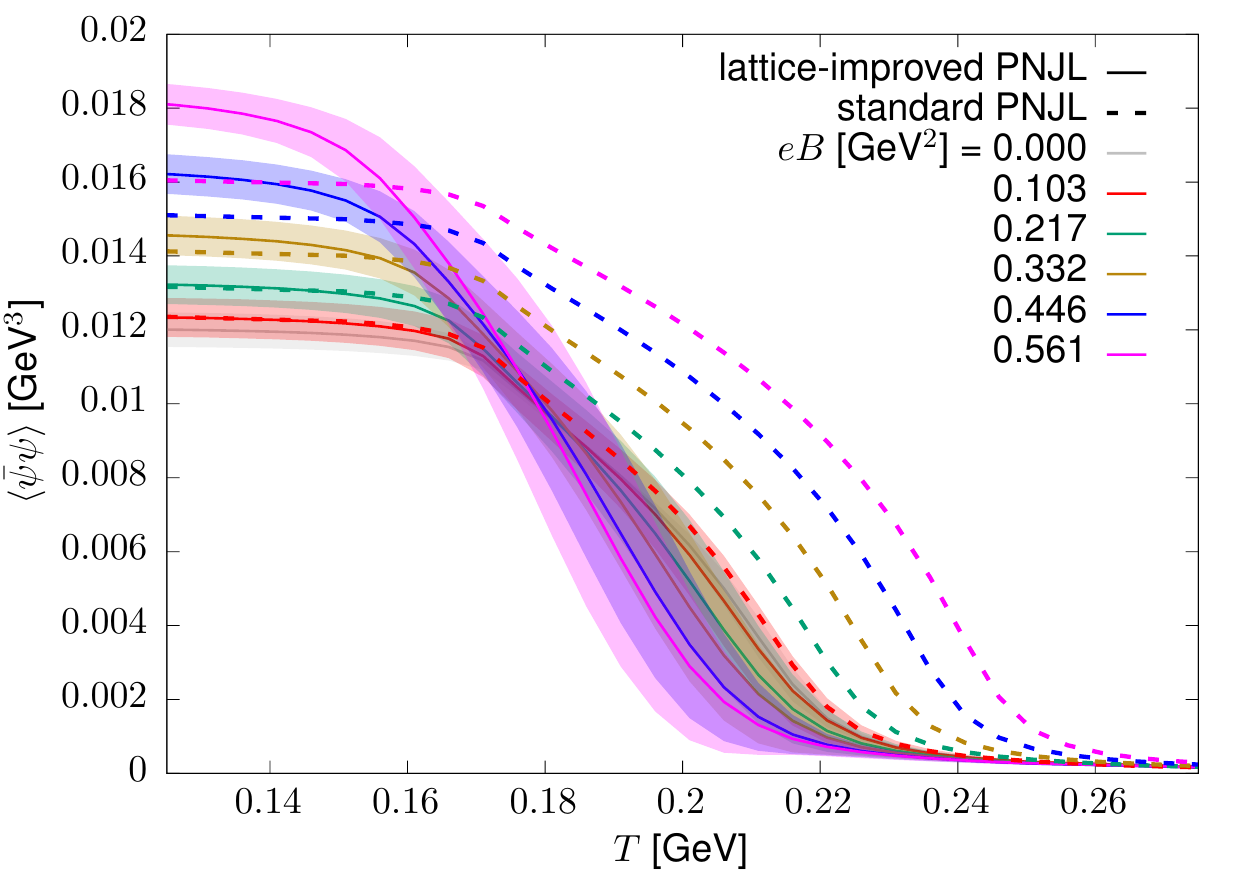}\hspace{0.05\textwidth}\includegraphics[width=0.475\textwidth]{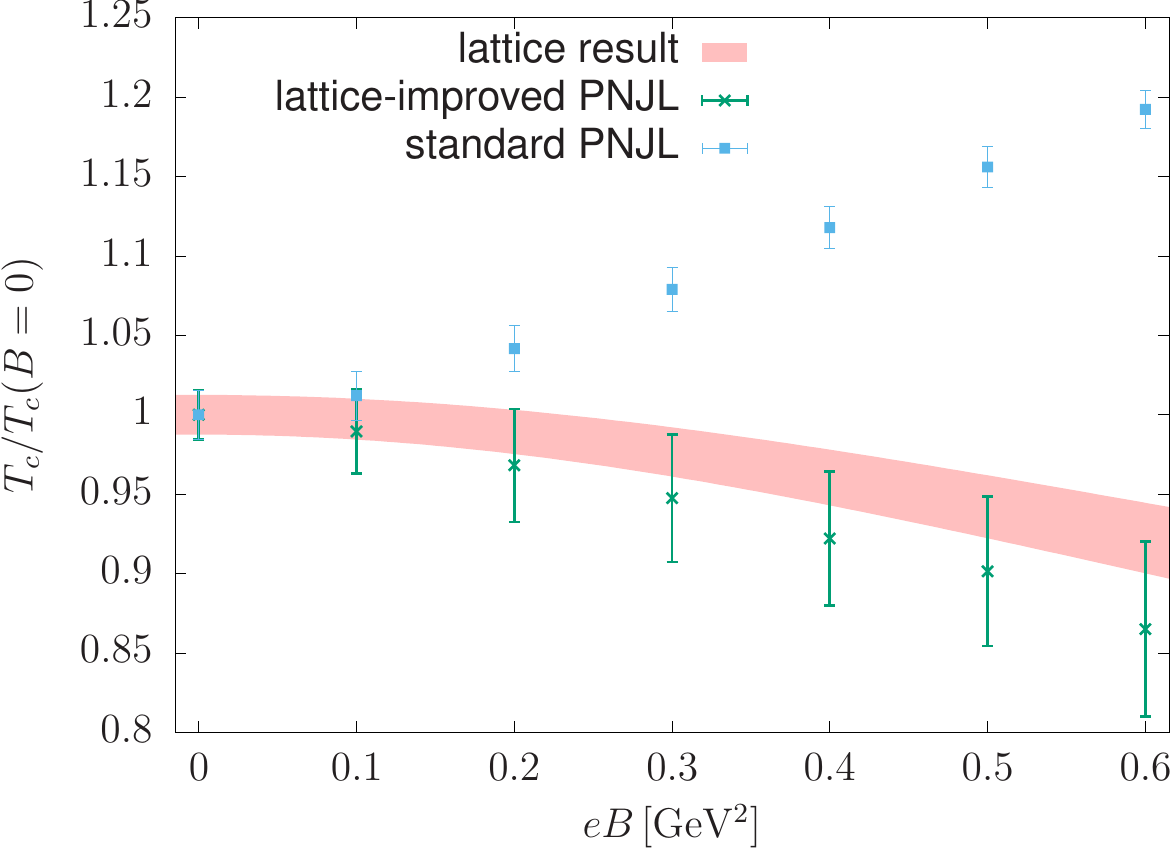}}\caption{\label{fig:pbp_n_Tc} Left: The quark condensate as a function of the temperature for different values of the magnetic field along the solutions shown in Fig.~\ref{fig:njlsols} compared with the standard PNJL results (for which errors are omitted again). While the lattice-improved PNJL model displays inverse magnetic catalysis the standard one misses this feature. Right: The pseudo-critical temperature as a function of the magnetic field from lattice simulations, the lattice-improved PNJL and the standard PNJL model scaled by their respective $B=0$ values.}
\end{figure}
\end{center}

We define two pseudo-critical temperatures: the inflection point of the quark condensate, which is identified with the chiral transition temperature, and the inflection point of the Polyakov loop, which in turn is typically associated to the deconfinement transition, however we will only discuss in detail the one obtained from the quark condensate now. In the left panel of Fig.~\ref{fig:pbp_n_Tc} we show the quark condensate curves corresponding to the solutions shown in Fig.~\ref{fig:njlsols}, which display inverse magnetic catalysis around the transition in the case of the lattice-improved PNJL model, however not in the standard PNJL model, where magnetic catalysis can be seen at all temperatures. 
According to our results, the transition remains an analytic crossover 
for all magnetic fields under consideration, just as the lattice studies
found~\cite{Bali:2011qj,Endrodi:2015oba}. This is in contrast to the 
Polyakov loop-extended quark meson model, where a $B$-dependent tuning 
of model parameters was observed to induce a first-order phase transition already 
at low $B$~\cite{Fraga:2013ova}.

In the right panel of Fig.~\ref{fig:njlsols} we compare the $T_c(B)$ curves with lattice results from Ref.~\cite{Bali:2011qj} where we see that after rescaling with the respective $T_c(B=0)$ values the lattice-improved result is consistent with the lattice continuum limit as opposed to the standard PNJL result. The pseudo-critical temperature at vanishing magnetic field in the lattice-improved PNJL model is $T_c(B=0) = 204(3)$ MeV. The deconfinement temperature defined from the Polyakov-loop seems to be lower (similar behavior was found in \cite{Gatto:2010pt}) but the disentanglement of the two transitions may need more in-depth analysis.

\section{Summary}
\label{sec:summary}

In this paper we performed the first lattice determination of the
baryon spectrum in the presence of strong magnetic fields $B$ at the physical point, including a continuum extrapolation.
Using the $B$-dependence of the nucleon and $\Sigma$ baryon masses 
and assuming a simplistic quark model,
we defined constituent quark masses that were employed as zero-temperature inputs for the Polyakov loop-extended NJL model. 
The standard variant of this model is known to qualitatively fail in describing the QCD 
phase diagram in the magnetic field-temperature plane. 
We demonstrate that our {\it lattice-improved} PNJL model reproduces 
all features of the lattice findings at $B>0$, including the 
inverse magnetic catalysis of the light quark condensate in the 
transition region as well as the reduction of the chiral crossover temperature 
by $B$. This result reveals that the model can be substantially 
improved if minimal information is fed to it at zero temperature -- 
allowing it to capture the non-trivial dependence of $\bar\psi\psi(B,T)$ in a broad range of magnetic 
fields and temperatures.

An obvious extension of our results would be to include the 
strange quark flavor in the PNJL model or isospin splittings as well 
as further channels that may emerge at $B>0$~\cite{Ferrer:2013noa}.
The ideas presented in this work might also be generalized 
to other low-energy models of QCD.

\acknowledgments
This research was supported by the DFG (Emmy Noether Programme EN 1064/2-1), the Tempus Public Foundation under contract no.\ M\'AE\"O2018-2019/280643 and this work is part of Project No. 121064 for which support was provided by the National Research, Development and Innovation Fund of Hungary. The authors are grateful to Eduardo Fraga, S\'andor Katz, Sebastian Schmalzbauer, Igor Shovkovy and Zsolt Sz\'ep for enlightening discussions.

\bibliographystyle{jhep_new}
\bibliography{njlB}

\providecommand{\href}[2]{#2}\begingroup\raggedright\begin{thebibliography}{10}

\bibitem{Kharzeev:2015znc}
D.~E. Kharzeev, J.~Liao, S.~A. Voloshin and G.~Wang, \emph{{Chiral magnetic and
  vortical effects in high-energy nuclear collisions -- A status report}},
  \href{http://dx.doi.org/10.1016/j.ppnp.2016.01.001}{\emph{Prog. Part. Nucl.
  Phys.} {\bf 88} (2016) 1--28}, [\href{http://arxiv.org/abs/1511.04050}{{\tt
  1511.04050}}].

\bibitem{Giunti:2014ixa}
C.~Giunti and A.~Studenikin, \emph{{Neutrino electromagnetic interactions: a
  window to new physics}},
  \href{http://dx.doi.org/10.1103/RevModPhys.87.531}{\emph{Rev. Mod. Phys.}
  {\bf 87} (2015) 531}, [\href{http://arxiv.org/abs/1403.6344}{{\tt
  1403.6344}}].

\bibitem{Voronyuk:2011jd}
V.~Voronyuk, V.~D. Toneev, W.~Cassing, E.~L. Bratkovskaya, V.~P. Konchakovski
  and S.~A. Voloshin, \emph{{(Electro-)Magnetic field evolution in relativistic
  heavy-ion collisions}},
  \href{http://dx.doi.org/10.1103/PhysRevC.83.054911}{\emph{Phys. Rev.} {\bf
  C83} (2011) 054911}, [\href{http://arxiv.org/abs/1103.4239}{{\tt
  1103.4239}}].

\bibitem{Tuchin:2013ie}
K.~Tuchin, \emph{{Particle production in strong electromagnetic fields in
  relativistic heavy-ion collisions}},
  \href{http://dx.doi.org/10.1155/2013/490495}{\emph{Adv. High Energy Phys.}
  {\bf 2013} (2013) 490495}, [\href{http://arxiv.org/abs/1301.0099}{{\tt
  1301.0099}}].

\bibitem{Chernodub:2010qx}
M.~N. Chernodub, \emph{{Superconductivity of QCD vacuum in strong magnetic
  field}}, \href{http://dx.doi.org/10.1103/PhysRevD.82.085011}{\emph{Phys.
  Rev.} {\bf D82} (2010) 085011}, [\href{http://arxiv.org/abs/1008.1055}{{\tt
  1008.1055}}].

\bibitem{Andersen:2014xxa}
J.~O. Andersen, W.~R. Naylor and A.~Tranberg, \emph{{Phase diagram of QCD in a
  magnetic field: A review}},
  \href{http://dx.doi.org/10.1103/RevModPhys.88.025001}{\emph{Rev. Mod. Phys.}
  {\bf 88} (2016) 025001}, [\href{http://arxiv.org/abs/1411.7176}{{\tt
  1411.7176}}].

\bibitem{Aoki:2006we}
Y.~Aoki, G.~Endr\H{o}di, Z.~Fodor, S.~D. Katz and K.~K. Szab\'o, \emph{{The
  Order of the quantum chromodynamics transition predicted by the standard
  model of particle physics}},
  \href{http://dx.doi.org/10.1038/nature05120}{\emph{Nature} {\bf 443} (2006)
  675--678}, [\href{http://arxiv.org/abs/hep-lat/0611014}{{\tt
  hep-lat/0611014}}].

\bibitem{Bhattacharya:2014ara}
T.~Bhattacharya, M.~I. Buchoff, N.~H. Christ, H.-T. Ding, R.~Gupta et~al.,
  \emph{{QCD Phase Transition with Chiral Quarks and Physical Quark Masses}},
  \href{http://dx.doi.org/10.1103/PhysRevLett.113.082001}{\emph{Phys.Rev.Lett.}
  {\bf 113} (2014) 082001}, [\href{http://arxiv.org/abs/1402.5175}{{\tt
  1402.5175}}].

\bibitem{DElia:2010abb}
M.~D'Elia, S.~Mukherjee and F.~Sanfilippo, \emph{{QCD Phase Transition in a
  Strong Magnetic Background}},
  \href{http://dx.doi.org/10.1103/PhysRevD.82.051501}{\emph{Phys. Rev.} {\bf
  D82} (2010) 051501}, [\href{http://arxiv.org/abs/1005.5365}{{\tt
  1005.5365}}].

\bibitem{Bali:2011qj}
G.~S. Bali, F.~Bruckmann, G.~Endr\H{o}di, Z.~Fodor, S.~D. Katz, S.~Krieg
  et~al., \emph{{The QCD phase diagram for external magnetic fields}},
  \href{http://dx.doi.org/10.1007/JHEP02(2012)044}{\emph{JHEP} {\bf 02} (2012)
  044}, [\href{http://arxiv.org/abs/1111.4956}{{\tt 1111.4956}}].

\bibitem{Bali:2012zg}
G.~S. Bali, F.~Bruckmann, G.~Endr\H{o}di, Z.~Fodor, S.~D. Katz and
  A.~Sch{\"a}fer, \emph{{QCD quark condensate in external magnetic fields}},
  \href{http://dx.doi.org/10.1103/PhysRevD.86.071502}{\emph{Phys. Rev.} {\bf
  D86} (2012) 071502}, [\href{http://arxiv.org/abs/1206.4205}{{\tt
  1206.4205}}].

\bibitem{Endrodi:2015oba}
G.~Endr\H{o}di, \emph{{Critical point in the QCD phase diagram for extremely
  strong background magnetic fields}},
  \href{http://dx.doi.org/10.1007/JHEP07(2015)173}{\emph{JHEP} {\bf 07} (2015)
  173}, [\href{http://arxiv.org/abs/1504.08280}{{\tt 1504.08280}}].

\bibitem{Shovkovy:2012zn}
I.~A. Shovkovy, \emph{{Magnetic Catalysis: A Review}},
  \href{http://dx.doi.org/10.1007/978-3-642-37305-3_2}{\emph{Lect. Notes Phys.}
  {\bf 871} (2013) 13--49}, [\href{http://arxiv.org/abs/1207.5081}{{\tt
  1207.5081}}].

\bibitem{Bruckmann:2013oba}
F.~Bruckmann, G.~Endr\H{o}di and T.~G. Kov\'acs, \emph{{Inverse magnetic
  catalysis and the Polyakov loop}},
  \href{http://dx.doi.org/10.1007/JHEP04(2013)112}{\emph{JHEP} {\bf 04} (2013)
  112}, [\href{http://arxiv.org/abs/1303.3972}{{\tt 1303.3972}}].

\bibitem{Gusynin:1995nb}
V.~P. Gusynin, V.~A. Miransky and I.~A. Shovkovy, \emph{{Dimensional reduction
  and catalysis of dynamical symmetry breaking by a magnetic field}},
  \href{http://dx.doi.org/10.1016/0550-3213(96)00021-1}{\emph{Nucl. Phys.} {\bf
  B462} (1996) 249--290}, [\href{http://arxiv.org/abs/hep-ph/9509320}{{\tt
  hep-ph/9509320}}].

\bibitem{Bruckmann:2017pft}
F.~Bruckmann, G.~Endr\H{o}di, M.~Giordano, S.~D. Katz, T.~G. Kov\'acs,
  F.~Pittler et~al., \emph{{Landau levels in QCD}},
  \href{http://dx.doi.org/10.1103/PhysRevD.96.074506}{\emph{Phys. Rev.} {\bf
  D96} (2017) 074506}, [\href{http://arxiv.org/abs/1705.10210}{{\tt
  1705.10210}}].

\bibitem{Endrodi:2019zrl}
G.~Endr\H{o}di, M.~Giordano, S.~D. Katz, T.~G. Kov\'acs and F.~Pittler,
  \emph{{Magnetic catalysis and inverse catalysis for heavy pions}},
  \href{http://arxiv.org/abs/1904.10296}{{\tt 1904.10296}}.

\bibitem{DElia:2018xwo}
M.~D'Elia, F.~Manigrasso, F.~Negro and F.~Sanfilippo, \emph{{QCD phase diagram
  in a magnetic background for different values of the pion mass}},
  \href{http://dx.doi.org/10.1103/PhysRevD.98.054509}{\emph{Phys. Rev.} {\bf
  D98} (2018) 054509}, [\href{http://arxiv.org/abs/1808.07008}{{\tt
  1808.07008}}].

\bibitem{Fraga:2008qn}
E.~S. Fraga and A.~J. Mizher, \emph{{Chiral transition in a strong magnetic
  background}}, \href{http://dx.doi.org/10.1103/PhysRevD.78.025016}{\emph{Phys.
  Rev.} {\bf D78} (2008) 025016}, [\href{http://arxiv.org/abs/0804.1452}{{\tt
  0804.1452}}].

\bibitem{Gatto:2010pt}
R.~Gatto and M.~Ruggieri, \emph{{Deconfinement and Chiral Symmetry Restoration
  in a Strong Magnetic Background}},
  \href{http://dx.doi.org/10.1103/PhysRevD.83.034016}{\emph{Phys. Rev.} {\bf
  D83} (2011) 034016}, [\href{http://arxiv.org/abs/1012.1291}{{\tt
  1012.1291}}].

\bibitem{Fraga:2012fs}
E.~S. Fraga and L.~F. Palhares, \emph{{Deconfinement in the presence of a
  strong magnetic background: an exercise within the MIT bag model}},
  \href{http://dx.doi.org/10.1103/PhysRevD.86.016008}{\emph{Phys. Rev.} {\bf
  D86} (2012) 016008}, [\href{http://arxiv.org/abs/1201.5881}{{\tt
  1201.5881}}].

\bibitem{Fraga:2012ev}
E.~S. Fraga, J.~Noronha and L.~F. Palhares, \emph{{Large $N_c$ Deconfinement
  Transition in the Presence of a Magnetic Field}},
  \href{http://dx.doi.org/10.1103/PhysRevD.87.114014}{\emph{Phys. Rev.} {\bf
  D87} (2013) 114014}, [\href{http://arxiv.org/abs/1207.7094}{{\tt
  1207.7094}}].

\bibitem{Fraga:2012rr}
E.~S. Fraga, \emph{{Thermal chiral and deconfining transitions in the presence
  of a magnetic background}},
  \href{http://dx.doi.org/10.1007/978-3-642-37305-3_5}{\emph{Lect. Notes Phys.}
  {\bf 871} (2013) 121--141}, [\href{http://arxiv.org/abs/1208.0917}{{\tt
  1208.0917}}].

\bibitem{Fraga:2013ova}
E.~S. Fraga, B.~W. Mintz and J.~Schaffner-Bielich, \emph{{A search for inverse
  magnetic catalysis in thermal quark-meson models}},
  \href{http://dx.doi.org/10.1016/j.physletb.2014.02.028}{\emph{Phys. Lett.}
  {\bf B731} (2014) 154--158}, [\href{http://arxiv.org/abs/1311.3964}{{\tt
  1311.3964}}].

\bibitem{Farias:2014eca}
R.~L.~S. Farias, K.~P. Gomes, G.~I. Krein and M.~B. Pinto, \emph{{Importance of
  asymptotic freedom for the pseudocritical temperature in magnetized quark
  matter}}, \href{http://dx.doi.org/10.1103/PhysRevC.90.025203}{\emph{Phys.
  Rev.} {\bf C90} (2014) 025203}, [\href{http://arxiv.org/abs/1404.3931}{{\tt
  1404.3931}}].

\bibitem{Ferreira:2014kpa}
M.~Ferreira, P.~Costa, O.~Lourenco, T.~Frederico and C.~Providencia,
  \emph{{Inverse magnetic catalysis in the (2+1)-flavor Nambu-Jona-Lasinio and
  Polyakov-Nambu-Jona-Lasinio models}},
  \href{http://dx.doi.org/10.1103/PhysRevD.89.116011}{\emph{Phys. Rev.} {\bf
  D89} (2014) 116011}, [\href{http://arxiv.org/abs/1404.5577}{{\tt
  1404.5577}}].

\bibitem{Ayala:2014iba}
A.~Ayala, M.~Loewe, A.~J. Mizher and R.~Zamora, \emph{{Inverse magnetic
  catalysis for the chiral transition induced by thermo-magnetic effects on the
  coupling constant}},
  \href{http://dx.doi.org/10.1103/PhysRevD.90.036001}{\emph{Phys. Rev.} {\bf
  D90} (2014) 036001}, [\href{http://arxiv.org/abs/1406.3885}{{\tt
  1406.3885}}].

\bibitem{Ayala:2014gwa}
A.~Ayala, M.~Loewe and R.~Zamora, \emph{{Inverse magnetic catalysis in the
  linear sigma model with quarks}},
  \href{http://dx.doi.org/10.1103/PhysRevD.91.016002}{\emph{Phys. Rev.} {\bf
  D91} (2015) 016002}, [\href{http://arxiv.org/abs/1406.7408}{{\tt
  1406.7408}}].

\bibitem{Ferreira:2013tba}
M.~Ferreira, P.~Costa, D.~P. Menezes, C.~Providência and N.~Scoccola,
  \emph{{Deconfinement and chiral restoration within the SU(3)
  Polyakov--Nambu--Jona-Lasinio and entangled Polyakov--Nambu--Jona-Lasinio
  models in an external magnetic field}},
  \href{http://dx.doi.org/10.1103/PhysRevD.89.016002,
  10.1103/PhysRevD.89.019902}{\emph{Phys. Rev.} {\bf D89} (2014) 016002},
  [\href{http://arxiv.org/abs/1305.4751}{{\tt 1305.4751}}].

\bibitem{Braun:2014fua}
J.~Braun, W.~A. Mian and S.~Rechenberger, \emph{{Delayed Magnetic Catalysis}},
  \href{http://dx.doi.org/10.1016/j.physletb.2016.02.017}{\emph{Phys. Lett.}
  {\bf B755} (2016) 265--269}, [\href{http://arxiv.org/abs/1412.6025}{{\tt
  1412.6025}}].

\bibitem{Andersen:2014oaa}
J.~O. Andersen, W.~R. Naylor and A.~Tranberg, \emph{{Inverse magnetic catalysis
  and regularization in the quark-meson model}},
  \href{http://dx.doi.org/10.1007/JHEP02(2015)042}{\emph{JHEP} {\bf 02} (2015)
  042}, [\href{http://arxiv.org/abs/1410.5247}{{\tt 1410.5247}}].

\bibitem{Mueller:2015fka}
N.~M{\"u}ller and J.~M. Pawlowski, \emph{{Magnetic catalysis and inverse
  magnetic catalysis in QCD}},
  \href{http://dx.doi.org/10.1103/PhysRevD.91.116010}{\emph{Phys. Rev.} {\bf
  D91} (2015) 116010}, [\href{http://arxiv.org/abs/1502.08011}{{\tt
  1502.08011}}].

\bibitem{Avancini:2016fgq}
S.~S. Avancini, R.~L.~S. Farias, M.~Benghi~Pinto, W.~R. Tavares and V.~S.
  Timóteo, \emph{{$\pi_0$ pole mass calculation in a strong magnetic field and
  lattice constraints}},
  \href{http://dx.doi.org/10.1016/j.physletb.2017.02.002}{\emph{Phys. Lett.}
  {\bf B767} (2017) 247--252}, [\href{http://arxiv.org/abs/1606.05754}{{\tt
  1606.05754}}].

\bibitem{Farias:2016gmy}
R.~L.~S. Farias, V.~S. Timoteo, S.~S. Avancini, M.~B. Pinto and G.~Krein,
  \emph{{Thermo-magnetic effects in quark matter: Nambu--Jona-Lasinio model
  constrained by lattice QCD}},
  \href{http://dx.doi.org/10.1140/epja/i2017-12320-8}{\emph{Eur. Phys. J.} {\bf
  A53} (2017) 101}, [\href{http://arxiv.org/abs/1603.03847}{{\tt 1603.03847}}].

\bibitem{Brandt:2017oyy}
B.~B. Brandt, G.~Endr\H{o}di and S.~Schmalzbauer, \emph{{QCD phase diagram for
  nonzero isospin-asymmetry}},
  \href{http://dx.doi.org/10.1103/PhysRevD.97.054514}{\emph{Phys. Rev.} {\bf
  D97} (2018) 054514}, [\href{http://arxiv.org/abs/1712.08190}{{\tt
  1712.08190}}].

\bibitem{Braghin:2016zba}
F.~L. Braghin, \emph{{SU(2) low energy quark effective couplings in weak
  external magnetic field}},
  \href{http://dx.doi.org/10.1103/PhysRevD.94.074030}{\emph{Phys. Rev.} {\bf
  D94} (2016) 074030}, [\href{http://arxiv.org/abs/1606.05587}{{\tt
  1606.05587}}].

\bibitem{Braghin:2017zas}
F.~L. Braghin, \emph{{Low energy constituent quark and pion effective couplings
  in a weak external magnetic field}},
  \href{http://dx.doi.org/10.1140/epja/i2018-12485-6}{\emph{Eur. Phys. J.} {\bf
  A54} (2018) 45}, [\href{http://arxiv.org/abs/1705.05926}{{\tt 1705.05926}}].

\bibitem{Aarts:2018glk}
G.~Aarts, C.~Allton, D.~De~Boni and B.~Jäger, \emph{{Hyperons in thermal QCD:
  A lattice view}},
  \href{http://dx.doi.org/10.1103/PhysRevD.99.074503}{\emph{Phys. Rev.} {\bf
  D99} (2019) 074503}, [\href{http://arxiv.org/abs/1812.07393}{{\tt
  1812.07393}}].

\bibitem{Aarts:2017rrl}
G.~Aarts, C.~Allton, D.~De~Boni, S.~Hands, B.~Jäger, C.~Praki et~al.,
  \emph{{Light baryons below and above the deconfinement transition: medium
  effects and parity doubling}},
  \href{http://dx.doi.org/10.1007/JHEP06(2017)034}{\emph{JHEP} {\bf 06} (2017)
  034}, [\href{http://arxiv.org/abs/1703.09246}{{\tt 1703.09246}}].

\bibitem{AliKhan:1993qk}
A.~Ali~Khan, M.~Gockeler, R.~Horsley, P.~E.~L. Rakow, G.~Schierholz and
  H.~Stuben, \emph{{Spectroscopy and renormalization group flow of a lattice
  Nambu-Jona-Lasinio model}},
  \href{http://dx.doi.org/10.1103/PhysRevD.51.3751}{\emph{Phys. Rev.} {\bf D51}
  (1995) 3751--3780}, [\href{http://arxiv.org/abs/hep-lat/9401012}{{\tt
  hep-lat/9401012}}].

\bibitem{Martinelli:1982cb}
G.~Martinelli, G.~Parisi, R.~Petronzio and F.~Rapuano, \emph{{The Proton and
  Neutron Magnetic Moments in Lattice {QCD}}},
  \href{http://dx.doi.org/10.1016/0370-2693(82)90162-9}{\emph{Phys. Lett.} {\bf
  116B} (1982) 434--436}.

\bibitem{Chang:2015qxa}
{\scshape NPLQCD} collaboration, E.~Chang, W.~Detmold, K.~Orginos, A.~Parreno,
  M.~J. Savage, B.~C. Tiburzi et~al., \emph{{Magnetic structure of light nuclei
  from lattice QCD}},
  \href{http://dx.doi.org/10.1103/PhysRevD.92.114502}{\emph{Phys. Rev.} {\bf
  D92} (2015) 114502}, [\href{http://arxiv.org/abs/1506.05518}{{\tt
  1506.05518}}].

\bibitem{Parreno:2016fwu}
A.~Parreno, M.~J. Savage, B.~C. Tiburzi, J.~Wilhelm, E.~Chang, W.~Detmold
  et~al., \emph{{Octet baryon magnetic moments from lattice QCD: Approaching
  experiment from a three-flavor symmetric point}},
  \href{http://dx.doi.org/10.1103/PhysRevD.95.114513}{\emph{Phys. Rev.} {\bf
  D95} (2017) 114513}, [\href{http://arxiv.org/abs/1609.03985}{{\tt
  1609.03985}}].

\bibitem{Hidaka:2012mz}
Y.~Hidaka and A.~Yamamoto, \emph{{Charged vector mesons in a strong magnetic
  field}}, \href{http://dx.doi.org/10.1103/PhysRevD.87.094502}{\emph{Phys.
  Rev.} {\bf D87} (2013) 094502}, [\href{http://arxiv.org/abs/1209.0007}{{\tt
  1209.0007}}].

\bibitem{Bali:2017ian}
G.~S. Bali, B.~B. Brandt, G.~Endr\H{o}di and B.~Gl{\"a}{\ss}le, \emph{{Meson
  masses in electromagnetic fields with Wilson fermions}},
  \href{http://dx.doi.org/10.1103/PhysRevD.97.034505}{\emph{Phys. Rev.} {\bf
  D97} (2018) 034505}, [\href{http://arxiv.org/abs/1707.05600}{{\tt
  1707.05600}}].

\bibitem{Bali:2018sey}
G.~S. Bali, B.~B. Brandt, G.~Endr\H{o}di and B.~Gl{\"a}{\ss}le, \emph{{Weak
  decay of magnetized pions}},
  \href{http://dx.doi.org/10.1103/PhysRevLett.121.072001}{\emph{Phys. Rev.
  Lett.} {\bf 121} (2018) 072001}, [\href{http://arxiv.org/abs/1805.10971}{{\tt
  1805.10971}}].

\bibitem{Bonati:2015dka}
C.~Bonati, M.~D'Elia and A.~Rucci, \emph{{Heavy quarkonia in strong magnetic
  fields}}, \href{http://dx.doi.org/10.1103/PhysRevD.92.054014}{\emph{Phys.
  Rev.} {\bf D92} (2015) 054014}, [\href{http://arxiv.org/abs/1506.07890}{{\tt
  1506.07890}}].

\bibitem{Aoki:2005vt}
Y.~Aoki, Z.~Fodor, S.~D. Katz and K.~K. Szab\'o, \emph{{The Equation of state
  in lattice QCD: With physical quark masses towards the continuum limit}},
  \href{http://dx.doi.org/10.1088/1126-6708/2006/01/089}{\emph{JHEP} {\bf 01}
  (2006) 089}, [\href{http://arxiv.org/abs/hep-lat/0510084}{{\tt
  hep-lat/0510084}}].

\bibitem{Borsanyi:2010cj}
S.~Bors\'anyi et~al., \emph{{The QCD equation of state with dynamical quarks}},
  \href{http://dx.doi.org/10.1007/JHEP11(2010)077}{\emph{JHEP} {\bf 11} (2010)
  077}, [\href{http://arxiv.org/abs/1007.2580}{{\tt 1007.2580}}].

\bibitem{Ishizuka:1993mt}
N.~Ishizuka, M.~Fukugita, H.~Mino, M.~Okawa and A.~Ukawa, \emph{{Operator
  dependence of hadron masses for Kogut-Susskind quarks on the lattice}},
  \href{http://dx.doi.org/10.1016/0550-3213(94)90475-8}{\emph{Nucl. Phys.} {\bf
  B411} (1994) 875--902}.

\bibitem{Durr:2008zz}
S.~D{\"urr} et~al., \emph{{Ab-Initio Determination of Light Hadron Masses}},
  \href{http://dx.doi.org/10.1126/science.1163233}{\emph{Science} {\bf 322}
  (2008) 1224--1227}, [\href{http://arxiv.org/abs/0906.3599}{{\tt 0906.3599}}].

\bibitem{Taya:2014nha}
H.~Taya, \emph{{Hadron Masses in Strong Magnetic Fields}},
  \href{http://dx.doi.org/10.1103/PhysRevD.92.014038}{\emph{Phys. Rev.} {\bf
  D92} (2015) 014038}, [\href{http://arxiv.org/abs/1412.6877}{{\tt
  1412.6877}}].

\bibitem{Fukushima:2010fe}
K.~Fukushima, M.~Ruggieri and R.~Gatto, \emph{{Chiral magnetic effect in the
  PNJL model}}, \href{http://dx.doi.org/10.1103/PhysRevD.81.114031}{\emph{Phys.
  Rev.} {\bf D81} (2010) 114031}, [\href{http://arxiv.org/abs/1003.0047}{{\tt
  1003.0047}}].

\bibitem{Klevansky:1992qe}
S.~P. Klevansky, \emph{{The Nambu-Jona-Lasinio model of quantum
  chromodynamics}},
  \href{http://dx.doi.org/10.1103/RevModPhys.64.649}{\emph{Rev. Mod. Phys.}
  {\bf 64} (1992) 649--708}.

\bibitem{Ratti:2005jh}
C.~Ratti, M.~A. Thaler and W.~Weise, \emph{{Phases of QCD: Lattice
  thermodynamics and a field theoretical model}},
  \href{http://dx.doi.org/10.1103/PhysRevD.73.014019}{\emph{Phys. Rev.} {\bf
  D73} (2006) 014019}, [\href{http://arxiv.org/abs/hep-ph/0506234}{{\tt
  hep-ph/0506234}}].

\bibitem{Schaefer:2007pw}
B.-J. Schaefer, J.~M. Pawlowski and J.~Wambach, \emph{{The Phase Structure of
  the Polyakov--Quark-Meson Model}},
  \href{http://dx.doi.org/10.1103/PhysRevD.76.074023}{\emph{Phys. Rev.} {\bf
  D76} (2007) 074023}, [\href{http://arxiv.org/abs/0704.3234}{{\tt
  0704.3234}}].

\bibitem{Avancini:2019wed}
S.~S. Avancini, R.~L.~S. Farias, N.~N. Scoccola and W.~R. Tavares,
  \emph{{NJL-type models in the presence of intense magnetic fields: the role
  of the regularization prescription}},
  \href{http://arxiv.org/abs/1904.02730}{{\tt 1904.02730}}.

\bibitem{Cohen:2007bt}
T.~D. Cohen, D.~A. McGady and E.~S. Werbos, \emph{{The Chiral condensate in a
  constant electromagnetic field}},
  \href{http://dx.doi.org/10.1103/PhysRevC.76.055201}{\emph{Phys. Rev.} {\bf
  C76} (2007) 055201}, [\href{http://arxiv.org/abs/0706.3208}{{\tt
  0706.3208}}].

\bibitem{Ferrer:2013noa}
E.~J. Ferrer, V.~de~la Incera, I.~Portillo and M.~Quiroz, \emph{{New look at
  the QCD ground state in a magnetic field}},
  \href{http://dx.doi.org/10.1103/PhysRevD.89.085034}{\emph{Phys. Rev.} {\bf
  D89} (2014) 085034}, [\href{http://arxiv.org/abs/1311.3400}{{\tt
  1311.3400}}].

\end{thebibliography}\endgroup

\end{document}